\documentclass[superscriptaddress,preprintnumbers,amsmath,amssymb,twocolumn,floats,aps,prl]{revtex4-1}
\usepackage{txfonts}
\usepackage{amssymb}
\usepackage{graphicx}
\usepackage{sidecap}
\usepackage{CJK}
\usepackage{txfonts}
\usepackage{url}

\begin{document}

\bibliographystyle{unsrt}
\title{Charge transfer effects in naturally occurring van der Waals heterostructures (PbSe)$_{1.16}$(TiSe$_2$)$_m$ ($m=1,~2$)}

\author{Q. Yao}
\affiliation{State Key Laboratory of Surface Physics, Department of Physics, and Laboratory of Advanced Materials, Fudan University, Shanghai 200433, China}
\affiliation{State Key Laboratory of Functional Materials for Informatics, Shanghai Institute of Microsystem and Information Technology (SIMIT), Chinese Academy of Sciences, Shanghai 200050, China}
\affiliation{Collaborative Innovation Centre of Advanced Microstructures, Nanjing 210093, China}

\author{D. W. Shen}
\email{dwshen@mail.sim.ac.cn}
\affiliation{State Key Laboratory of Functional Materials for Informatics, Shanghai Institute of Microsystem and Information Technology (SIMIT), Chinese Academy of Sciences, Shanghai 200050, China}
\affiliation{CAS Center for Excellence in Superconducting Electronics (CENSE), Shanghai 200050, China}

\author{C. H. P. Wen}
\affiliation{State Key Laboratory of Surface Physics, Department of Physics, and Laboratory of Advanced Materials, Fudan University, Shanghai 200433, China}
\affiliation{Collaborative Innovation Centre of Advanced Microstructures, Nanjing 210093, China}

\author{C. Q. Hua}
\affiliation{Department of Physics, Zhejiang University, Hangzhou 310027, China}

\author{L. Q. Zhang}
\affiliation{Collaborative Innovation Centre of Advanced Microstructures, Nanjing 210093, China}
\affiliation{National Laboratory of Solid State Microstructures, College of Physics, Nanjing University, Nanjing 210093, China}

\author{N. Z. Wang}
\affiliation{Hefei National Laboratory for Physical Sciences at Microscale and Department of Physics and Key Laboratory of Strongly-coupled Quantum Matter Physics, University of Science and Technology of China, Hefei 230026, China}

\author{X. H. Niu}
\author {Q. Y. Chen}
\affiliation{State Key Laboratory of Surface Physics, Department of Physics, and Laboratory of Advanced Materials, Fudan University, Shanghai 200433, China}
\affiliation{Collaborative Innovation Centre of Advanced Microstructures, Nanjing 210093, China}

\author{P. Dudin}
\affiliation{Diamond Light Source, Harwell Science and Innovation Campus, Didcot OX11 0DE, United Kingdom}

\author{Y. H. Lu}
\affiliation{State Key Lab of Silicon Materials, Zhejiang University, Hangzhou 310027, China}

\author{Y. Zheng}
\affiliation{Department of Physics, Zhejiang University, Hangzhou 310027, China}

\author{X. H. Chen}
\affiliation{Collaborative Innovation Centre of Advanced Microstructures, Nanjing 210093, China}
\affiliation{Hefei National Laboratory for Physical Sciences at Microscale and Department of Physics and Key Laboratory of Strongly-coupled Quantum Matter Physics, University of Science and Technology of China, Hefei 230026, China}
\affiliation{High Magnetic Field Laboratory, Chinese Academy of Sciences, Hefei 230031, China}

\author{X. G. Wan}
\affiliation{Collaborative Innovation Centre of Advanced Microstructures, Nanjing 210093, China}
\affiliation{National Laboratory of Solid State Microstructures, College of Physics, Nanjing University, Nanjing 210093, China}

\author{D. L. Feng}
\affiliation{State Key Laboratory of Surface Physics, Department of Physics, and Laboratory of Advanced Materials, Fudan University, Shanghai 200433, China}
\affiliation{Collaborative Innovation Centre of Advanced Microstructures, Nanjing 210093, China}

\begin{abstract}
van der Waals heterostructures (vdWHs) exhibit rich properties and thus potential for  applications, and charge transfer between different layers in a heterostructure often dominates its properties and device performance. It is thus critical to reveal and understand the charge transfer effects in vdWHs, for which electronic structure measurements have proven to be effective. Using angle-resolved photoemission spectroscopy, we studied the electronic structures of (PbSe)$_{1.16}$(TiSe$_2$)$_m$ ($m=1,~2$), which are naturally occurring vdWHs, and discovered several striking charge transfer effects. When the thickness of the TiSe$_2$ layers is halved from $m=2$ to $m=1$, the amount of charge transfered increases unexpectedly by more than 250~\%. This is accompanied by a dramatic drop in the electron-phonon interaction strength far beyond the prediction by first-principles calculations and, consequently, superconductivity only exists in the $m=2$ compound with strong electron-phonon interaction, albeit with lower carrier density. Furthermore, we found that the amount of charge transfered in both compounds is nearly halved when warmed from below 10~K to room temperature, due to the different thermal expansion coefficients of the constituent layers of these misfit compounds. These unprecedentedly large charge transfer effects might widely exist in vdWHs composed of metal-semiconductor contacts, thus our results provide important insights for further understanding and applications of vdWHs.
\end{abstract}

\maketitle

Boosted by the enormous effort on graphene-like materials, research into artificial materials formed by stacking different two-dimensional (2D) crystals in a desired sequence, the so-called van der Waals heterostructures (vdWHs) have gradually drawn considerable attention~\cite{VDWHReview,VDWH1,VDWH2,VDWH3,VDWH4,VDWH5}. The interaction between neighboring 2D crystals is relatively weak, but owing to the atomically flat interfaces and precise crystallographic alignment, electronic orbitals can still protrude across the interfaces and thus affect charge carriers in the adjacent 2D crystals. The resulting charge reconstruction leads to emergent properties distinct from those of its individual components, {\it e.g.}, the observation of Hofstadter's butterfly~\cite{Hofstadter_Butterfly1, Hofstadter_Butterfly2}, the fractal quantum hall effect~\cite{quantum_Hall_effect1, quantum_Hall_effect2} and the unconventional diode and photovoltaic effects in MoS$_2$/WSe$_2$ heterostructures~\cite{diode_effect_and_photovoltaic_effect}, which naturally pique both fundamental and technological interest. However, the deterministic placement methods for preparing most artificial vdWHs impede the experimental investigation of the influence of charge transfer on their low-energy electronic structure, which is the key to understanding the underlying mechanism and of important directive significance to further devices based on vdWHs.

Misfit compounds, bulk materials which consist of alternately-stacked rock-salt and transition metal chalcogenide layers, belong to naturally occurring vdWHs~\cite{misfit1,misfit2,misfit3,VDWH6,VDWH7}. The neighboring structural subsystems exhibit different symmetry and periodicity, leading to an incommensurate crystal structure bound by weak van der Waals interactions between interleaved layers, as sketched in Fig. 1(a)~\cite{misfit_review}. These compounds provide a unique opportunity to investigate the manipulation of vdWHs, and in particular the cleavable crystal structure and absence of unwanted interlayer adsorbates enable the direct probing of their intrinsic electronic structure by angle-resolved photoemission spectroscopy (ARPES).

In this Letter, through systematic ARPES measurements and first-principles calculations on the low-lying electronic structure of typical naturally occurring vdWHs (PbSe)$_{1.16}$(TiSe$_2$)$_m$ ($m=1,~2$) (denoted as m1 and m2 hereinafter), we demonstrate remarkable charge transfer effects which may prove endemic to vdWHs. Firstly, varying the thickness of TiSe$_2$ layers from m2 to m1 results in the significant increase in the charge transfered at the interfaces, which is accompanied by a dramatic drop in the electron-phonon coupling (EPC) strength far beyond the prediction by first-principles calculations. Furthermore, an unexpected temperature-dependent band shift is revealed, which introduces an up to 40~\% decrease in the charge transfer when warmed from 10\,K to room temperature. Since a large number of current prototype vdWH devices are made from transition metal chalcogenides, and their design is optimized mainly based on their low-temperature properties, this finding may prove crucial to the further exploration of vdWHs applications.


\begin{figure}[t]
\includegraphics[width=\columnwidth]{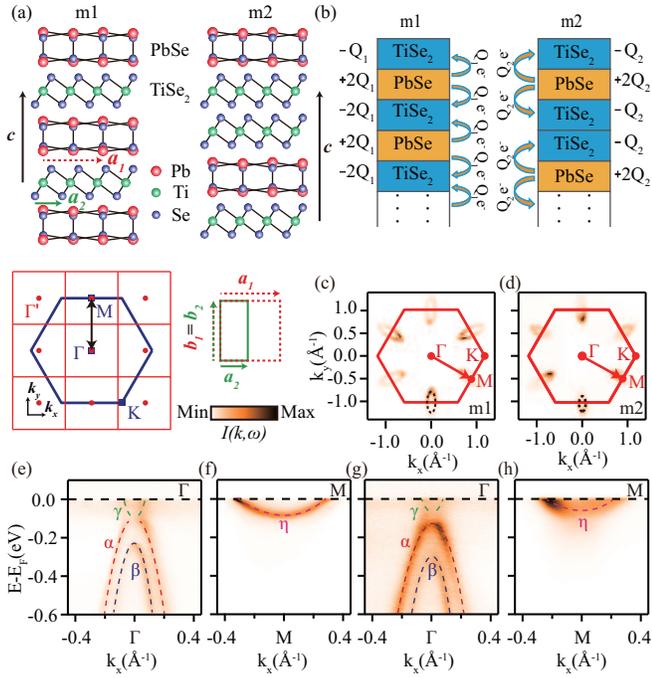}
\caption{(a) Crystal structure of (PbSe)$_{1.16}$(TiSe$_2$)$_m$ ($m=1,~2$). The bottom left depicts the projected 2D Brillouin zones (BZ) of the TiSe$_2$ (blue hexagon) and PbSe layers (red squares). Note that the $\Gamma$ point of the second PbSe BZ and the M point of the TiSe$_2$ layer coincide. The bottom right shows a comparison of their lattice constants. (b) Schematic diagram of the charge transfer at interfaces. Orange and cyan rectangles represent PbSe and TiSe$_2$ layers, respectively. PbSe layers donate electrons to TiSe$_2$ layers. (c) and (d) Photoemission intensity maps of m1 and m2 at $E_F$, respectively, over the projected 2D BZ. Black dashed ellipses indicate the Fermi pockets and the intensity has been integrated over a window of ($E_F-2.5$\,meV, $E_F+2.5$\,meV). Data were threefold symmetrized. Photoemission intensity plots (e) and (f) for m1 and (g) and (h) for m2 are presented near \emph{$\Gamma$} and \emph{M}, respectively. The dashed lines indicate band dispersions obtained from the momentum and energy distribution curves (MDCs and EDCs). All data were taken at 10~K with 80~eV photons.}
\label{crystal structure}
\end{figure}

\begin{figure}[b]
\includegraphics[width=\columnwidth]{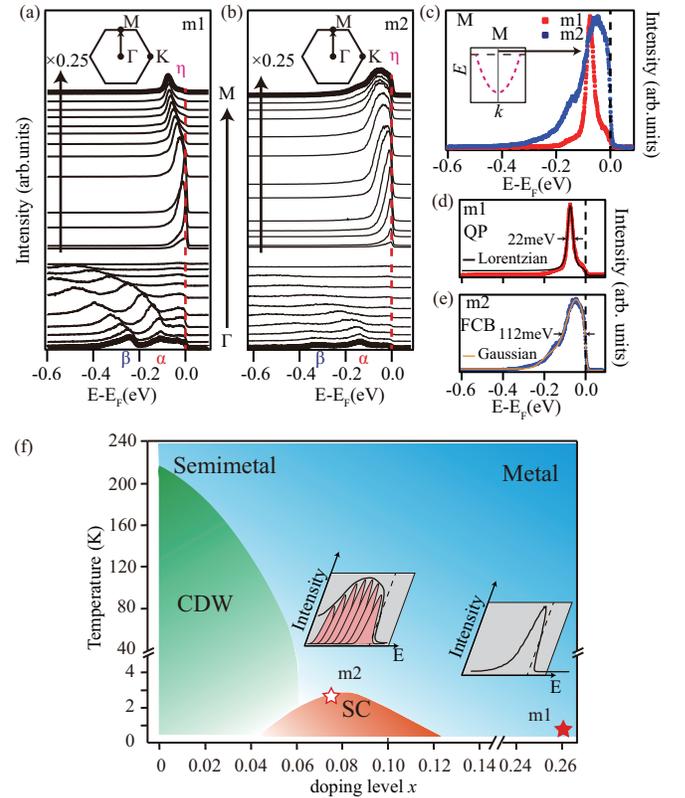}
\caption{(a) and (b) EDCs along the \emph{$\Gamma$}-\emph{M} cuts shown in Fig. 1(c) and 1(d) for m1 and m2, respectively. Spectra around \emph{M} have been intendedly scaled down by 0.25 to be comparable to those around \emph{$\Gamma$}. (c) Comparison of their EDCs at the \emph{M} point. The lineshape (d) for m1 can be fitted by a Lorentzian peak, while that (e) for m2 is fit best by a Gaussian envelope. (f) The phase diagram of electron doped TiSe$_2$, whose regime below 0.1 doping is  reproduced from that of Cu$_x$TiSe$_2$ with one Cu ion doping one electron ~\cite{CuTiSe2_nature}, and the overdoped regime is a sketch. m1 is located in the heavily overdoped regime with quasiparticle-like spectra, which extends the original phase diagram, while m2 is located in the optimally-doped regime and its spectra exhibit the typical Franck-Condon broadening of polarons due to strong EPC.}
\label{Polaronic superconductivity}
\end{figure}

High-quality (PbSe)$_{1.16}$(TiSe$_2$)$_m$ ($m=1,~2$) single crystals were synthesized by a chemical vapor transport method as described elsewhere~\cite{Chen_Xianhui}. They could be mechanically exfoliated down to few-layer thickness, forming natural and air-stable vdWHs suitable for device applications (see Supplemental Material~\cite{SM} for details). ARPES measurements were performed at Beamline 5-4 of Stanford Synchrotron Radiation Laboratory (SSRL), Beamline I05-ARPES of Diamond Light Source and Beamline 13U of National Synchrotron Radiation Laboratory (NSRL). These beamlines are all equipped with VG-Scienta R4000 electron analyzers. The overall energy resolution was 10$\sim$15~meV depending on the photon energy, and the angular resolution was set to 0.3$^\circ$. All samples were cleaved under a vacuum better than 5$\times$10$^{-11}$ Torr. The details of first-principles calculations can be found in Supplemental Material~\cite{SM, SM1, SM2, SM3, SM4}.

\begin{figure*}
    \includegraphics[width=18cm]{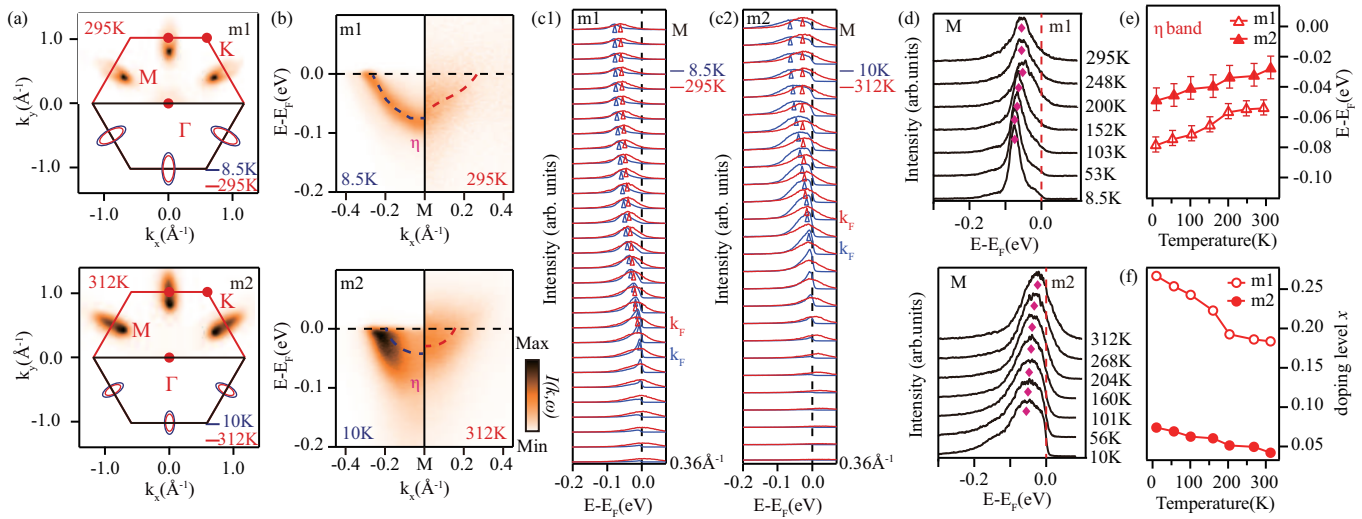}
    \caption{(a) Fermi surfaces of (PbSe)$_{1.16}$(TiSe$_2$)$_m$ ($m=1,~2$). The upper halves are room-temperature photoemission intensity maps of m1 and m2 integrated over ($E_F-2.5$\,meV, $E_F+2.5$\,meV). Data were sixfold symmetrized. The lower halves are a quantitative comparison of Fermi pocket size at low (8.5 and 10\,K) and room (295 and 312\,K) temperatures. (b) Comparisons of photoemission intensity plots around \emph{M} for m1 and m2 at low and ambient temperatures. (c1-c2) Direct comparisons of the EDCs taken at high and low temperatures for m1 and m2, respectively. (d) Temperature-dependent EDCs located at \emph{M} for m1 and m2, respectively. (e) Temperature dependence of the $\eta$ band bottom energy for m1 and m2. (f) Temperature dependence of carrier concentrations for m1 and m2 due to charge transfer.}
\label{temperature-dependence Fermi surface and dispersion}
\end{figure*}

We first present the photoemission intensity maps of m1 [Fig.~\ref{crystal structure}(c)] and m2 [Fig.~\ref{crystal structure}(d)] at the Fermi energy $E_F$. Both full-scale maps show hexagonal symmetry in accord with the lattice of bulk $1T$-TiSe$_2$. According to the crystal structure of misfit compounds, there should be sizable possibility for PbSe layer to be present on the surface. However, neither map shows any fourfold symmetric feature from the PbSe layers in all of our more than a dozen samples. We speculate that the random or incommensurate potential may affect PbSe  severely and localize its holes, and its photoemission signal  is thus smeared in momentum space.

Both Fermi surface maps exhibit elliptical Fermi pockets around the $M$ point similar to bulk $1T$-TiSe$_2$ ~\cite{bulk_TiSe2}, but a detailed comparison indicates significant electron doping in both m1 and m2. The estimated Luttinger volumes of the Fermi pockets are 13\% and 7.4\% of the Brillouin zone for m1 and m2, respectively (see more details in Supplemental Material~\cite{SM}). Based on the charge transfer processes sketched in Fig.~\ref{crystal structure}(b), the doped charge in the surface layer of m1 would be half its bulk value, while there is no charge reconstruction in m2, which gives bulk doping levels of 0.26~$e^-$ and 0.074~$e^-$ per TiSe$_2$ for m1 and m2, respectively~\cite{Luttinger, polar_catastrophe, xiaohai, pengrui}. This finding confirms the previous model that the PbSe layers serve as a carrier reservoir for TiSe$_2$ layers, thus giving rise to the charge transfer. When the thickness of the TiSe$_2$ layers is halved from m2 to m1, the amount of transfered charge increases by as large as 250~\%, which suggests that varying the thickness of the interleaved layers might provide a rather efficient means of tuning charge transfer in vdWHs. We note that the charge transfer originates from the difference in Fermi level/chemical potential of neighboring layers in vdWHs~\cite{SM}. Although our findings are  on bulk misfit compounds, they represent the behaviors of vdWHs down to a few unit cells. As shown by  the static electric model analysis in Supplemental Material~\cite{SM,SM5}, the charge distribution is stable and independent of the number of unit cells down to 2 u.c. of PbSe/TiSe$_2$, 2 u.c. PbSe/TiSe$_2$/TiSe$_2$, or 1 u.c. TiSe$_2$/PbSe/TiSe$_2$.

More detailed electronic structure comparisons of m1 and m2 are shown in Figs. 1(e-h). In the vicinity of $E_F$, one electron-like ($\gamma$) and two hole-like bands ($\alpha$ and $\beta$) around $\varGamma$ together with another electron-like band ($\eta$) around \emph{M} can be clearly resolved. As in the bulk material, $\eta$ is a Ti 3$d$ band while the highly-dispersive $\alpha$ and $\beta$ are spin-orbit-split Se 4$p$ bands~\cite{Se4p_1,Se4p_2}. However, we stress that the $\gamma$ band in (PbSe)$_{1.16}$(TiSe$_2$)$_m$ cannot be assigned as scattered states by the charge-density-wave (CDW) wave vector as in $1T$-TiSe$_2$, since CDW is unlikely present in m1 and m2 for the following reasons. Firstly, throughout the whole temperature range from 300K to 10K, we have never found any sign of band folding in our ARPES measurements, which is  the smoking gun for the presence of bulk or surface CDW states in $1T$-TiSe$_2$~\cite{Zhao_Jiafeng}. Secondly, our ARPES measurements have confirmed nearly the same photoemission spectra of m2 as those of Cu$_{0.65}$TiSe$_2$, which is located in the same optimal doping regime as m2; CDW has been proven absent at this doping~\cite{Zhao_Jiafeng,CuTiSe2_nature}. Thirdly, the doping level of m1 is much higher than m2 on the surface, and even higher in the bulk, and thus m1 is further away from the CDW regime than m2. Instead, $\gamma$  is more reasonably accounted-for as the electronic states of TiSe$_2$ scattered from \emph{M} by the reciprocal lattice constant of the PbSe layers [illustrated by the double-headed arrow in Fig. 1(a)]. This result suggests that the charge density from the PbSe layers impinges upon the TiSe$_2$ planes, serving as a commensurate potential that scatters the electrons therein~\cite{misfit_Ou}.

The dramatic change in the transfered charge with $m$ is accompanied by an unexpected modulation of the photoemission spectra of these naturally occurring vdWHs. Figs.~\ref{Polaronic superconductivity}(a) and \ref{Polaronic superconductivity}(b) compare EDCs of m1 and m2 along $\varGamma$-\emph{M} at low temperatures. Note that while the $\eta$ band of m1 shows typical quasiparticle-like dispersion with sharp and well-defined peaks, the spectral lineshape of m2 taken at the same momenta is remarkably broad and the linewidth is on the same order as the dispersion, which strongly suggests the incoherent nature of the spectra. Meanwhile, the maxima of these incoherent peaks still follows the dispersion predicted by band structure calculations. Taking spectra around the $\eta$ band bottoms of both compounds as an example [Fig.~\ref{Polaronic superconductivity}(c)], rather different behavior is clearly observed. The EDC for m1 is well fit by a Lorentzian peak having a linewidth of 22.1~meV, in which the shadow band background and the Fermi-Dirac function have been considered [Fig.~\ref{Polaronic superconductivity}(d)]. In contrast, as shown in Fig.~\ref{Polaronic superconductivity}(e), the lineshape of the spectrum taken from m2 is clearly non-quasiparticle-like, and it can only be reasonably fit by a Gaussian function with a linewidth of 112~meV, more than 5 times wider than m1. Note that both samples are rather 2D in nature (see Supplemental Material~\cite{SM} for details, ~\cite{SM7}) and transport measurements have further proven the more 2D electronic structure of m2. Furthermore, X-ray diffraction and transition electron microscopy have illustrated the similar crystalline quality of our m1 and m2 samples~\cite{Chen_Xianhui}. Therefore, this unconventional spectroscopic broadening of m2 is not due to $k_z$ broadening or disorder. Moreover, since the actual doping in bulk is twice of that on the surface for m1, its linewidth of bulk states should be even much smaller.

Such spectroscopic linewidth broadening has also been reported at low doping $x$ in Cu$_x$TiSe$_2$~\cite{Zhao_Jiafeng} --- photoemission spectra of Cu$_{0.065}$TiSe$_2$, located in the same optimal doping regime as m2, exhibit an analogous broad feature. The observed sharp Fermi edges in both our data and Ref.~\cite{Zhao_Jiafeng} represent the Fermi-Dirac cutoff of a broad feature extending well above $E_F$. This behavior is characteristic of the typical polarons in transition metal chalcogenides~\cite{polaron4, TMD_polaron}, indicating the very strong EPC in m2, which is reasonable as it is in proximity to a CDW phase~\cite{SunPRB}. In principle, our first-principles calculations can reproduce the tendency for diminishing EPC strength with increasing electron doping in TiSe$_2$, with the coupling constant $\lambda$ decreasing from 0.55 in m2 to 0.51 in m1 (see Supplemental Material~\cite{SM} for more details of calculations, ~\cite{SM8, SM9, SM10}). However, the $>$500\,\%\ difference in the spectroscopic linewidth between these two sibling vdWHs is not well accounted-for by this calculation, suggesting a much more drastic change in the EPC strength owing to charge transfer. Our findings also suggest that the strong EPC would likely reduce the charge transfer and cause the observed large difference between m1 and m2 --- the two TiSe$_2$ layers in m2 would share electrons transfered from one layer of PbSe, so that the carrier concentration in its TiSe$_2$ layer would be less than the m1 case, which results in stronger EPC that subsequently reduces the charge transfer. This possibility needs to be tested by further calculations including the doping dependent EPC. Moreover, one immediate consequence of such a significant modulation of the EPC is on the emergence of superconductivity in (PbSe)$_{1.16}$(TiSe$_2$)$_m$ ($m=1,~2$). As summarized in the phase diagram of electron doped TiSe$_2$ [Fig.~\ref{Polaronic superconductivity}(f)], the spectra of m2 exhibit typical Franck-Condon broadening of polarons due to rather strong EPC, while those of m1 behave like the coherent quasiparticle in a weak-coupling regime (see more analysis in Supplemental Material~\cite{SM}). The weak EPC in m1 explains its absence of superconductivity despite the higher density of states near $E_F$ than in m2.

\begin{figure}[t]
    \includegraphics[width=\columnwidth]{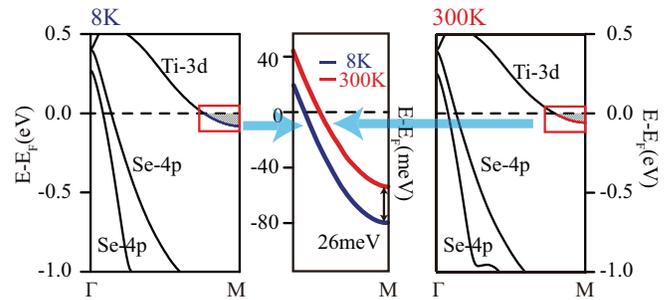}
    \caption{Band structure comparison of the TiSe$_2$ component in misfit vdWHs at low and room temperatures, as simulated by first-principles calculations based on the thermal expansion of the lattice.}
\label{summary}
\end{figure}

Unexpectedly, we discovered that the carrier concentrations of these naturally occurring vdWHs exhibit strong temperature-dependence. The upper halves of Fig.~\ref{temperature-dependence Fermi surface and dispersion}(a) show the room temperature photoemission intensity maps of m1 and m2. Compared with those taken at low temperatures, the electron pockets around \emph{M} shrink markedly. The more quantitative comparison of Fermi surface contours at different temperatures [the lower part of Fig.~\ref{temperature-dependence Fermi surface and dispersion}(a)], extracted from the MDC peak fitting, further confirms that the charge transfered from PbSe to TiSe$_2$ is reduced for both m1 and m2 on increasing temperature. This shrinking of the Fermi pockets can be attributed to the upward shift of the Ti $3d$ $\eta$ band [Fig.~\ref{temperature-dependence Fermi surface and dispersion}(b)], which could be directly observed from comparisons of EDCs taken at high and low temperatures for m1 and m2 as well [Figs.~\ref{temperature-dependence Fermi surface and dispersion}(c1-c2)]. Detailed temperature dependence measurements indicate that this band shift is continuous and not due to some sudden phase transition, as illustrated in Fig.~\ref{temperature-dependence Fermi surface and dispersion}(d). Quantitatively, the $\eta$ bands of m1 and m2 shift as much as 24.5 and 21.2~meV towards $E_F$, respectively, from our lowest measurement temperature to room temperature [Fig.~\ref{temperature-dependence Fermi surface and dispersion}(e)].

To account for the temperature-dependent energy shift of the $\eta$ band, we have considered the thermal expansion effect of the TiSe$_2$ lattice and then estimated the Ti $3d$ band shift through band structure calculations (Fig.~\ref{summary})~\cite{TiSe2_lattice}. This band energy shift is calculated to be about 26~meV from 8 to 300\,K, in good agreement with our experimental results (see Supplemental Material~\cite{SM} for details of this calculation). Since the chemical potential of PbSe lays higher in energy than the Fermi level of TiSe$_2$~\cite{Materials} and the upward shift of Ti $3d$ band observed on warming in misfit vdWHs would reduce the energy gap between them, the electron transfer from PbSe to TiSe$_2$ will be effectively suppressed. Accordingly, the resulting carrier concentrations of m1 and m2 are modulated by as much as 30~\% and 40~\%, respectively, by temperature [Fig.~\ref{temperature-dependence Fermi surface and dispersion}(f)]. Currently, most prototype devices based on vdWHs are designed mainly according to their low-temperature properties, but are intended to be employed in a room temperature environment. If such a huge variation in carrier concentration with temperature occurs more generally in vdWHs, it is essential that it should be taken into account.


To summarize, we have comprehensively characterized the electronic structure of the typical naturally occurring vdWHs (PbSe)$_{1.16}$(TiSe$_2$)$_m$ ($m=1,~2$). By varying the thickness of the TiSe$_2$ layers, we can realize carrier concentration modulation of these vdWHs on an unprecedentedly large scale. Simultaneously, the EPC strength in these vdWHs can be tuned {\it via} charge transfer, and thus the dome-like dependence of the superconducting transition temperature with carrier concentration in electron-doped TiSe$_2$ can be understood. Most remarkably, a dramatic temperature-dependent band shift is revealed which cuts the carrier concentration nearly in half on warming to room temperature. These findings may provide crucial feedback for the application of devices based on vdWHs and offer a new direction to further our understanding of this interesting class of materials.

We acknowledge Diamond Light Source for time on beamline I05 under proposal SI14737, which contributed to the results presented here. We gratefully acknowledge experimental support from Drs.\ D.\ H.\ Lu, T.\ K.\ Kim, M.\ Hoesch, L. G. Ma and helpful discussions with Drs.\ T.\ P.\ Ying, R.\ Peng and D.\ C.\ Peets. This work is supported by the National Key R\&D Program of the MOST of China (Grant No.\ 2016YFA0300200) and the National Science Foundation of China (Grants Nos.\ 11574337, 11227902 and U1332209). D.W.S. is also supported by ``Award for Outstanding Member in Youth Innovation Promotion Association CAS".

\bibliographystyle{apsrev4-1}

%

\end{document}